\documentstyle[pra,aps,twocolumn]{revtex}
\begin{document}
\draft

\newcommand{\bea}{\begin{array}}
\newcommand{\eea}{\end{array}}
\newcommand{\be}{\begin{equation}}
\newcommand{\ee}{\end{equation}}
\newcommand{\bm}{\begin{array}{cc}}
\newcommand{\emx}{\end{array}}
\newcommand{\bex}{\begin{eqnarray}}
\newcommand{\enx}{\end{eqnarray}}
\newcommand{\ben}{\begin{enumerate}}
\newcommand{\enn}{\end{enumerate}}
\newcommand{\bei}{\begin{itemize}}
\newcommand{\eei}{\end{itemize}}
\newcommand{\enxn}{\nonumber\end{eqnarray}}
\def\hcal{{\cal H}}
\def\ra{\rangle}
\def\la{\langle}
\def\dim{\rm dim}

\title{On balance of information in bipartite quantum communication
systems: entanglement-energy analogy}

\author{Ryszard Horodecki$^{1,}$\cite{poczta1},
Micha\l{} Horodecki$^{1,}$\cite{poczta2} and Pawe\l{} Horodecki
$^{2,}$\cite{poczta3}}

\address{$^1$ Institute of Theoretical Physics and Astrophysics,
University of Gda\'nsk, 80--952 Gda\'nsk, Poland,\\
$^2$Faculty of Applied Physics and Mathematics,
Technical University of Gda\'nsk, 80--952 Gda\'nsk, Poland
}

%

\def\hcal{{\cal H}}
\def\ecal{{\cal E}}
\def\ccal{{\cal C}}
\def\qcal{{\cal Q}}
\def\pcal{{\cal P}}

\def\dim{{\rm dim}}
\maketitle

\begin{abstract}
We adopt the view according to which information is the primary physical
entity that posseses objective meaning.
Basing on two postulates that (i) entanglement is a form of quantum
information corresponding to internal energy (ii) sending qubits corresponds
to work, we show that in the closed bipartite quantum communication systems
the information is conserved. We also discuss entanglement-energy analogy in
context of the Gibbs-Hemholtz-like equation connecting
the entanglement of formation, distillable entanglement and bound
entanglement. Then we show that in the deterministic protocols of
distillation the information is conserved.
We also discuss the objectivity of quantum information
in context of information interpretation of quantum states
and alghoritmic complexity.
\end{abstract}



\section{Introduction}
It is astonishing that just lately after over sixteen years quantum formalizm
reveals us new possibilities due to entanglement processing being a root of
new quantum phenomena such as quantum cryptography with Bell theorem
\cite{Ekert}, quantum dense coding \cite{geste},
quantum teleportation \cite{telep}, quantum computation \cite{Deutsh}.
It shows how important is to recognize
not only the structure of the formalizm itself but also potential possibilites
encoded in it.

In spite of many beautiful experimental and theoretical results on entanglement there
are still difficulties in understanding its many faces. It seems to be a reflection
of basic difficulties in the interpretation of quantum formalizm as well as
quantum-clasical hybridism in our perception of Nature. To overcome the latter
the existence of unitary information field being a necessary
condition of {\it any} communication (or correlation) has been
postulated \cite{P87,P89,tata} as well as the information
interpretation of quantum wavefunction has been considered \cite{P89}.
It rests on the generic information paradigm according to which the notion of
information represents a basic category and it can be defined independently of
probability itself \cite{Kolmogorow,IngardenU,IngardenK,tata}.
It implies that Nature is unbroken entity. However,
according to double, hylemorphic nature of the unitary information
field, there are two  mutually coupled levels of physical reality in Nature: logical
(informational) due to potential field of alternatives and energetic
due the field of activities (events)
\cite{foot1}.
Then from the point of view of the generic information paradigm, quantum
formalism is simply a set of extremely useful informational algorithms
describing the above complementary aspects of the same,
really existing unitary information field. It leads  in a natural  way to analogy
between information (entanglement) and energy being nothing but a reflection
of unity  of Nature.

Following this route, one attempts to find some useful
analogies in the quantum communication domain. Namely, physicists believe that
there should exist the laws governing entanglement processing in quantum
communication systems, that are analogous to those in thermodynamics.

Short history of this view has its origin in the papers by Bennett
{\it et al.} who announced a possible irreversibility  of the entanglement
distillation process \cite{Bennett96,conc}. Popescu and Rohrlich
\cite{PRohlich} have pointed out analogy between distillation-formation of
pure entangled states and Carnot cycle, and they have shown that entanglement
is extensive quantity. The authors formulated principle of entanglement
processing analogous
to the second principle of thermodynamics: ``{\it Entanglement cannot increase
under local quantum operations and classical communication}''.
Vedral and
Plenio \cite{VP} have considered the principle in detail and pointed out that
there is some (although not complete) analogy between efficiency of
distillation and efficiency
of Carnot cycle. In Refs \cite{MRH,APS} entanglement-energy analogy has been
developed and conservation of information in closed quantum systems has
been postulated in analogy with the first principle of thermodynamics:
{\it Entanglement of compound system does not change under unitary processes
on one of the subsystems}\cite{MRH}.
Then an attempt to formulate the counterpart of the second principle
in a way consistent with the above principle has been done  (since in
the original
Popescu-Rohrlich formulation entanglement was not conserved).
%
%

The main purpose of the paper is to support entanglement-energy
analogy by demonstration that in the closed bipartite quantum
communication system the information is conserved.
The paper is organised as follows. In section II we describe closed
quantum communication bipartite system.
The next section contains formal description of
balance of quantum information involving notions of physical
and logical work. In section IV we introduce the concept of useful
logical work in quantum communication. In next section we present
balance of information in teleportation. In section VI we discuss
entanglement-analogy in the context of the Gibbs-Hemholtz-like equation
connecting entanglement of formation, distillable entanglement
and bound entanglement. In section \ref{sec-dist} we present
the  balance of information in the process of distillation.
In the last section
we discuss the objectivity of quantum information
in context of information interpretation of quantum states
and alghoritmic complexisity..

\section{Closed quantum communication system: the model}%
Consider closed quantum communication (QC) system $U$ composite of
system $S$, measuring system $M$ and environment $R$
\be
U=S+M+R
\ee
where each system is split into Alice and Bob parts  $S_X,M_X,R_X$;
$X=A,B$.

It is assumed that Alice and Bob can control the system
$S_X$ which does not interact with environment $R_X$. The $M_X$ system
consists of $m_X$ qubits  and cotinuously interacts with environment $R_X$.
In result the system  $M_X$, palying the role
of ``ancilla'', is measured  in distinguished basis
$|x_1x_2\ldots\ra$, $x_i=0,1$ \cite{Zurek}.
In this sense the measurement is understand
here as {\it the process of irreversible entanglement
with some environment} and the system $R_{X}$ is to
ensure this ireversibility.
Note that in the above approach  the
evolution of the system is unitary i. e.
abandon the von Neumann projection postulate
which leads to violation of energy-momentum conservation
\cite{Pearle}.
Then acting on one part of entangled system, we have no way to {\it annihilate}
entanglement. The latter can change only by means of interacting of the
{\it both} entangled subsystems.
It may be objected that we can destroy entanglement e.g. by randomizing the
relative phases on the subsystems of interest. However, if the reduction of
wave packet is {\it not} regarded to be a real physical process, then the
above operation must be considered as entangling the subsystem with
some other system by means of a {\it unitary} transformation.
Then the entanglement will not vanish but it will {\it spread} over
all the three subsystems.

The operations Alice and Bob can perform in our QC system are:
\bei
\item Quantum communication: Alice and Bob  can exchange particles  from  the
system $S_X$.
\item ``Classical communication'': Alice and Bob can exchange particles from
the system $M_X$
\eei
Note that the  number of qubits of the systems $S_A$ and $S_B$ can change
but the total number of qubits of the system $M$ is conserved (similarly for
$S$). Besides Alice and Bob can perform  unitary transformation over the
system $M_X+S_X$, $X=A,B$.

We would like to stress one more that in our approach
the  measurement represents an irreversible entanglement
rather than the ``projection'' of the state.
To see it consider the case when Alice and Bob share
a singlet state and Alice performs a measurement on it.
The the initial state of the system $M_{A}+S_{A}+S_{B}$
($M_{A}$ represents the Alice's ancilla while  $S_{A}$, $S_{B}$
correspond to the particles forming a singlet state) is
\begin{eqnarray}
&&|\Psi\rangle_{M_{A}S_{A}S_{B}}=|0\rangle_{M_{A}}|\Psi_{A}^{singlet}\rangle= \nonumber \\
&&|0\rangle_{M_{A}}\frac{1}{\sqrt{2}}(|0\rangle_{S_{A}}| 1\rangle_{S_{B}}-
|1\rangle_{S_{A}}| 0\rangle_{S_{B}})
\end{eqnarray}
Then Alice performs the unitary operation $U$ on subsystem $M_{A}+S_{A}$.
This operation corresponds to the interaction between $M_{A}$ and $S_{A}$
and can be represented by C-NOT gate.
As a result the whole system is in the state
\begin{equation}
|\Psi'\rangle_{M_{A}S_{A}S_{B}}=\frac{1}{\sqrt{2}}
(|0\rangle_{M_{A}}|0\rangle_{S_{A}}| 1\rangle_{S_{B}}-
|1\rangle_{M_{A}}|1\rangle_{S_{A}}| 0\rangle_{S_{B}})
\end{equation}
Further $M_{A}$ can be irreversibly entangled with
environment system $R_{A}$ (which models the
irreversibility of the measurement).
But $R_{A}$ is still on Alice side, hence we have entanglement
between systems $(R_{A}+M_{A}+S_{A})$ and $S_{B}$ unchanged
and equal to $E=1$ e-bit.

Of course, there are some interpretational problems
if one imagines that Alice ``reads out''
the result of the measurement as then we encounter
problems coming from possible extension of the model by
the projection postulate. However that for practical
reasons (i. e. as far as quantum information qualitative
description is concerned) the informational
processes like e. g. quantum teleportation do not require reading the data.
Moreover, it must be noted that at the absence of the projection postulate
the above model can be veiwed as consistent with ``many worlds''
interpretation \cite{Everett}.

\section{Conservation of quantum information: formal description}
To determine balance of information in the closed system $U$ we adopt two basic
postulates \cite{MRH,APS}
\bei
\item[1.] Entanglement is a form of quantum information corresponding to
internal energy.
\item[2.] Sending qubits corresponds to work.
\eei

In accordance with  the  postulate 1, the information is physical quantity
that, in particular, should be {\it conserved} in closed
quantum systems, similarly as energy.
The second postulate allows to deal with communication {\it
processes} (in thermodynamics work is a functional of process).
To obtain the balance we must define our ``energy'' and ``work''
quantitatively. To this end consider system $X$ described in the Hilbert space
${\cal H}$, ${\rm dim} {\cal H}=d$ being in a state $\varrho_{X}$.
We define {\it informational content} $I_{X}$ of the state $\varrho_{X}$
as follows (cf. \cite{Barnett}):
\begin{equation}
I_{X}=\log {\rm dim } {\cal H} - S(\varrho_{X})
\label{wzor}
\end{equation}
where ${\rm dim} {\cal H}=d$, $S(\varrho_{X})\equiv S(X)$
are the dimension of the Hilbert space and the
von Neumann entropy of the system state.
Note that $I_{X}$ satisfies the inequality
$0=I_{X}^{min} \leq I_{X} \leq I_{X} \leq I_{X}^{max}=
\log {\rm dim } {\cal H}$
where $I_{X}^{min}$ and $I_{X}^{max}$
are the information content of the maximal mixed state
and pure state respectively.
Thus it is well defined quantity which measures informational
content of the system $\varrho_{X}$.

The formula (\ref{wzor}) needs some comment
as usually one interprets the von Neumann entropy as
a measure of information. In fact there is no contradiction.
Imagine for a moment that we admitt projection postulate
i. e. Alice knows the concrete result of the measurement.
Then the von Neumann entropy measures the information
gain {\it after} the measurement while the formula
(\ref{wzor}) corresponds to the information {\it prior}
the measurement and this information, in particular, is
maximal if the system is in pure state.
This is the reason while we use the name informational
{\it  content} as it has actual rather than potential
(i. e. related to the future measurement) character.
Below we shall see that, after we abandon the
projection postulate, the above formula allows to perform
a balance of quantum information in a consistent way.
Note that the Hilbert space dimension used in formula
(\ref{wzor}) is present also in definitions of
other notions (see below), in particular in the
case of useful logical work (sec. IV).
It plays, to some extent, the role similar to the
one in channels capacities theory or error correction
codes where dimension of ``error free'' subspace is a central notion.

Consider now the QC system $U$, being in
the initial pure state $\psi_{in}$, described by  general Alice-Bob Hilbert
space scheme as follows
\be
\left.\bea{c}
\hcal_A\\
\otimes\\
\hcal_{A'}
\eea
\quad\quad\otimes \quad\quad
\hcal_B\quad\right\}\quad \psi_{in},
\ee
where $\hcal_A\otimes\hcal_A'$, $\hcal_B$ are the Hilbert spaces of the
$S_A+M_A+R_A$ and $S_B+M_B+R_A$ respectively. Then
in accordance with (\ref{wzor}) the information contents
of the Alice and Bob  subsystems are defined as follows
\be
I_A=\log\dim(\hcal_A\otimes\hcal_{A'})-S(A+A');
\ee
\be
I_B=\log\dim\hcal_B-S(B),
\ee
where $\dim(\hcal_A\otimes\hcal_{A'})$ and $\dim \hcal_{B}$ are the dimensions
the corresponding Hilbert spaces while $S(A+A')$, $S(B)$ are the von Neumann
entropies of the subsystems.

Now, after  transmission of the system $A'$ to receiver (Bob) the Alice-Bob
Hilbert space scheme is given by
\be
\left.
\hcal_A
\quad\quad\otimes \quad\quad
\bea{c}
\hcal_B\\
\otimes\\
\hcal_{A'}
\eea \quad \right\}\quad \psi_{out}
\ee
and the total sytem $U$ is in the final state $\psi_{out}$.

Now, in accordance with  the above ``sending qubits -- work''
postulate we consider {\it physical work}  performed over the system $U$  being a
{\it physical transmission} of particles. Consequently, we define
$W_p$ as a number of sent qubits of the system $A'$
\be
W_p=\log\dim \hcal_{A'}.
\ee
Note that  after transmission of the system $A'$ to the Bob,
there is increase of the information content of his subsystem. Then we say
that the system $U$  performed  the {\it logical work} $W_l$ that is defined
as increase of the informational  content of the Bob (in general - receiver)
system.
\be
W_l=I^B_{out}-I^{B}_{in}
\ee
where  $I^B_{in}=I^B$, $I^B_{out}=I^{B+A'}$. Then one can regard the
physical work as sending ``matter'' while the logical work  --
sending ``form'' that is consistent with the assumed hylemorphic nature
of the information field.
Subsequently we can define initial  and final
entanglement of the system $U$ as
\be
E_{in}=S(B)=S(A+A');\quad E_{out}=S(A)=S(B+A'),
\ee
where obvious relations between the entropies of the subsystems hold.
Now, in accordace with the first  postalate,
 $E_{in}$ and $E_{out}$ are simply initial and final {\it potential}
informations contained in the total system. Having so defined quantities
it is not hard to obtain the following information balance equations
\be
E_{in}+W_p=E_{out}+W_l
\label{balance}
\ee
or equivalently
\be
I_{in}^A+I_{in}^B+2E_{in}= I_{out}^A+I_{out}^B+2E_{out}=const.
\ee
Note that the latter equation is compatible with the principle of
information conservation expressed in the
following form (equivalent to the one in the Introduction):
{\it For a compound quantum system a sum of information
contained in the subsystems and information
contained in entanglement is conserved in unitary processes} \cite{MRH}.

To see how the above formalism works,  consider two simple examples with ideal
quantum transmission.
Suppose,  Alice sends an {\it unentangled} qubit of the system  $S$  to Bob.
Then the  physical work  $W_p$ is equal  to 1 qubit. In
result the  informational content of Bob's system increases by 1, thus also
the logical work $W_l$ amounts to one qubit. Of course, in this case
both ``in'' and ``out'' entanglement are 0.

Suppose now that  Alice sends maximally {\it entangled} qubit to Bob. Here, again,
physical work is 1 qubit, and there is no initial entanglement.
However the final entanglement is one ebit  and logical work is 0, because
the state of the Bob system is now completely mixed.

Now we see that, according to the balance equation (\ref{balance}) the
difference $W_p-W_l$
between  the physical and logical work is  due to entanglement. Indeed,
as in the above example, sending particle may result in
increase of entanglement rather than performing nonzero logical work.

\section{Useful logical work: quantum communication}
The basic question arises in the context of quantum communication.  Does the
balance (\ref{balance}) distinguish between quantum and ``classical''
communication in our model? It follows from definition that the physical work
does not distinguish between these types of communication.  But what about
logical work? Suppose that Alice sent  to Bob a particle of the system $M_A$
in a pure state $|0\ra$. But in our model such state does not undergo
decoherence. Then the logical work $W_l$ is equal to one qubit.
Needless to say it is not quantum communication. Hence the logical work is
not ``useful'' in this case.

In quantum communication we are usually interested  in sending faithfully
any superpositions without decoherence. Therefore it is convenient to
introduce the notion of {\it useful logical work} as follows.

{\bf Definition. } Useful work is amount of qubits of the system $S$
transmitted without decoherence
\be
W_u=\log \dim \hcal,
\label{useful}
\ee
where $\hcal$ is the Hilbert space transmitted asymptotically faithfully.
The latter means that {\it any} state of this space would be transmitted
with asymptotically perfect fidelity. We see that the work performed
in previous example was not useful, since in result of the process,
only the states $|0\ra$ or $|1\ra$ can be transmitted faithfully.

\section{Balance of information in teleportation}
\begin{figure}
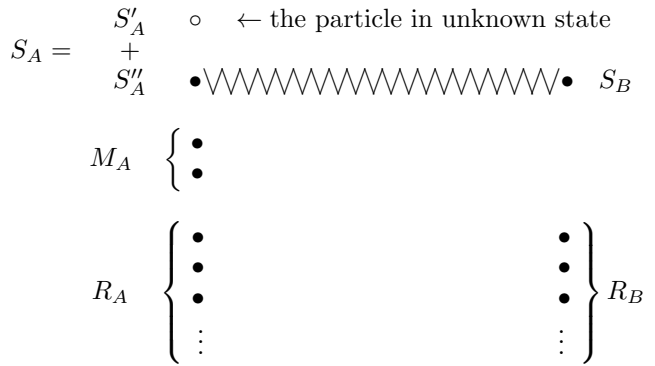

\def\strongpair{$\bullet\backslash\!/\!\backslash\!/\!\backslash\!/\!\backslash\!/%
\!\backslash\!/\!\backslash\!/\!\backslash\!/\!\backslash\!/\!\backslash\!/%
\!\backslash\!/\!\backslash\!/\!\backslash\!/\!\backslash\!/\!\backslash\!/%
\!\backslash\!/\!\backslash\!/\!\backslash\!/
\!\backslash\!/\!\backslash\!/\!%
\backslash\!/\bullet$}
\def\skoka{\hskip3mm}
\def\skok{\hskip4.7cm}
\[
\bea{l}
S_A= \quad
   \bea{c}
    S'_{A}\\
    +\\
    S''_{A}\\
   \eea
     \quad
   \bea{l}
    \circ\quad \leftarrow \mbox{the particle in unknown state}\\
    \\
    $\strongpair$\quad S_B\\
   \eea \\
\\
\hskip-3mm\skoka\quad\quad\quad M_A  \quad
\left\{
   \bea{c}
   \bullet\\
   \bullet\\
   \eea\right. \\
\\
\hskip-2.7mm\skoka\quad\quad\quad R_A\quad\left\{
   \bea{c}
   \bullet \skok \bullet\\
   \bullet \skok \bullet\\
   \bullet \skok \bullet\\
   \vdots \skok  \vdots\\
   \eea\right\} R_B\\
\eea
\]
\caption[model]{
\label{model}
The model of quantum communication system}
\end{figure}

To see how the above formalism works, consider the balance of quantum
information in teleportation \cite{telep}
\cite{foot2}.
Now the system $S_A$ consists
of a particle in unkown state and one particle from maximally entangled
pair, whereas the second particle from the pair represents $S_B$ system. The
system $M_A$ consists of two qubits  that interact with environment $R_A$
(Fig. \ref{model}).

The latter is only to ensure effective irreversibility of the
measurement and it is demonstrable that its action is irrelevant to the
information balance in the case of teleportation. As one knows, the initial
state can be written in the following form
\be
\psi_{in}\equiv\psi_0=\psi_{S'_A}^{unknown} \otimes \psi^{singlet}_{S_A''S_B}\otimes
|00\ra_{M_A},\quad
\ee
where $\psi_{S'_A}^{unknown}$ is the state
to be teleported, $\psi^{singlet}_{S_A''S_B}$ is the
singlet state of entangled pair and $|00\ra_{M_A}$ is the initial state
of the measuring system. It is easy to check that the initial entanglement $E_{in}$
of the initial state is equal to one $e-\mbox{bit}$. Now Alice performs ``measurement''
being local unitary transformation on her joint system $S_{A'}+S_{A''}+M_A$.
In result $\psi_{in}$ transforms to
\be
\psi_1={1\over 2}\sum_{i=0}^3 \psi^i_{S'_A S''_A}\otimes
\psi_B^{i(unknown)}\otimes |i\ra_{M_A},\quad
\ee
where $\psi_{S_A,S_{A''}}^i$ constitute Bell basis, $\psi_B^{i(unknown)}$
is rotated $\psi_{S_A}^{unknown}$,
$|i\ra_{M_A}$ is the state of the system $M_A$ indicating the result
of the measurement ($i$-th Bell state obtained). Since the Alice's operation is
unitary one, it does not change initial asymptotic entanglement. Subsequently,
Alice sends the two particles of the system $M_A$  to Bob. In accordance with
definition (6), it corresponds to two qubits $W_p=2$ of work performed over the
system. At the same time the state $\psi_1$ transforms to $\psi_2$ of the form
\be
\psi_2={1\over 2}\sum_{i=0}^3 \psi^i_{S'_A S''_A}\otimes
\psi_B^{i(unknown)}\otimes |i\ra_{M_B}.
\ee
Finally Bob decouples the system $S_B$ from other ones by unitary
transformation that of course does not change the asymptotic entanglement.

After classical communication from Alice entanglement of the total system
increased to the value $E_{out}=2$ e-bits.
Indeed Alice sends two particles of system $M_{A}$ to Bob
which are entangled with particles $S_{A}'$, $S_{A}''$).
On the other hand, the logical work performed by the system in the above
process amounts to $W_u=1$.
%
One can see the balance equation (\ref{balance}) is satisfied, and is of the
following form
\be
(E_{in}=1)+(W_p=2)=(E_{out}=2)+(W_u=1)
\ee
One easily recoginzes the result of the logical work in the
transmission of the unknown state to Bob. Since it is faithfully transmitted
independently of its particular form, we obtain that also useful logical
work  $W_u$ is equal to 1 qubit. Hence in the process of teleportation all
the work performed by the system is useful, and represents quantum
communication.

\section{Thermodynamical entanglement-energy analogy.
Gibbs-Hemholtz-like equation}
So far we have considered balance of information in closed
QC system. For open system (being, in general, in mixed
state) the situation is much more complicated being a reflection
of fundamental irreversibility in the asymptotic mixed state entanglement
processing \cite{Bennett96,conc,bound,irrev}.
Namely it has been shown \cite{bound} there is a
discontinuity in the structure of noisy entanglement.
It appeared that there are at least two quantitively
different types of entanglement: free - useful for quantum communication,
and bound - nondistillable, very weak and peculiar type of entanglement.
In accordance with  entanglement-energy analogy this new type of
entanglement was defined by equality
\begin{equation}
E_{F}=E_{bound}+E_{D},
\label{definicja}
\end{equation}
where $E_{F}$ and $E_{D}$ are asymptotic entanglement of formation
\cite{APS,footTot}
and distillable entanglement \cite{huge} respectively.
Note that for pure entangled states $|\Psi \rangle\langle \Psi|$ we have
always $E_{F}=E_{D}$, $E_{bound}=0$ \cite{conc}. Then, in this case
the whole entanglement can be converted into the useful quantum work
(see Fig. 2a)
with $E\equiv E_{F}(|\Psi \rangle\langle \Psi|)$).
For bound entangled mixed states we have $E_{D}=0$, $E_{F}=E_{bound}$.
It is quite likely that $E_{F}>0$ (so far we know only that
$E_{f}>0$ \cite{foot3}).
Here $E_{f}$ is entanglement of formation defined
in Ref. \cite{huge}.
 and then all prior nontrivial
entanglement of formation would be completely lost.
Thus in any process involving only separable or bound entangled
states useful quantum work is just zero.
In general, hovewer, it can happen that the state contains
two {\it different} types of entanglement.
Namely  there are  cases where $E_{bound}$ is strictly positive
i. e. we have \cite{irrev,footW}
\begin{equation}
E_{bound}=E_{F}-E_{D}>0.
\label{ostra}
\end{equation}
This reveals fundamental irreversibility in the domain
of quantum asymptotic information processing \cite{multi}.
It can be viewed as an analogue to irreversible thermodynamical processes
where only
the free energy (which is not equal to the total energy)
can be converted to useful work.
This supports the view \cite{APS}
according to which the equation (\ref{definicja})
can be regarded as quantum information counterpart
of the thermodynamical Gibbs-Hemholtz equation
$U=F+TS$ where quantities $E_{F}$, $E_{D}$,
$E_{bound}$ correspond to internal energy $U$, free energy $F$
and bound energy $TS$ respectively
($T$ and $S$ are the temperature and the entropy
of the system).


The above entanglement-energy analogy
has lead to the extension \cite{aktyw}
of the ``classical'' paradigm of LOCC operations by considering  new class
of entanglement processing called here entanglement enhanced LOCC
operations (EELOCC).
In particular, it suggested that entanglement can
be pumped from one to other system producing different
nonclasical chemical-like type processes.
In fact it allowed to find a new quantum effect - {\it activation} of bound
entanglement that corresponds to chemical activation process \cite{aktyw}.
Similarly, a recently discovered {\it cathalysis} of
pure entanglement involves EELOCC operations \cite{catal}.
In result the second principle of entanglement
processing (see Introduction) has been
generalized \cite{Lo} to cover the EELOCC paradigm:
{\it By local action, classical communication and N qubits of quantum
communication, entanglement cannot increase more than N e-bits.}


Now, it is interesting in the above context to consider
the problem of information balance in the cases
where systems are in mixed states.

\section{Balance of information in distillation process}
\label{sec-dist}
So far in our balance analysis the initial state of the  QC
system was pure. Let us consider the more general case. Suppose that
initial state of the system $S$ is mixed.
We have not generalized formalism to such case. We can however perform
balance of information in the case of the distillation process
\cite{Bennett96} (see in this context \cite{Rains}).
This task would be, in general, very difficult, because the almost all known
distillation protocols are {\it stochastic}. As one knows, the distillation
protocol aims at obtaining singlet pairs from a large amount of noisy pairs (in
mixed state) by LOCC operations. A convenient form of such a process would be
the following: Alice and Bob start with $n$ pairs, and after distillation
protocol, end up with $m$ singlet pairs. Such a protocol we shall call
{\it deterministic}.
Unfortunately, in the stochastic protocols
the situation is more complicated: Alice and Bob get with some probabilities
different number of output distilled pairs:
\[
\varrho_{in}=
\underbrace{\varrho\otimes\varrho\otimes\cdots\otimes \varrho}_n
\quad \rightarrow\quad
\left\{\bea{l}
\rightarrow p_0,\  \mbox{no output singlets}\\
\rightarrow p_1, \ \mbox{one output singlet}\\
\rightarrow p_2,\ \mbox{two output singlet}\\
\vdots\\
\eea\right.
\]
Since we must to describe the process in terms of closed system,
we will not see the above probabilities, but only their amplitudes.
As a result, we will have {\it no} clear distinction between the part
of the system containing distilled singlet pairs and the part containing
the remaining states of no useful entanglement.

Consider for example the first stage of the Bennett {\it et al.}
\cite{telep}
recursive protocol. It involves the folowing steps
\bei
\item take two two-spin 1/2 pairs, each in input state $\varrho$
\item perform operation $XOR\otimes XOR$
\item measure locally the spins  of  the target pair, and:
\bei
\item if the spins agree (probability $p_{a}$), keep the source pair
\item if the spins disagree (probability $p_{d}$), discard both pairs
\eei
\eei
After this operation we have the following final ``ensemble''
\[
\{(p_{a},\mbox{one pair in a new state $\tilde\varrho$}),\ \
(p_{d}, \mbox{no pairs})\}
\]
If we include environment  to the description, the events ``no pair''
and ``one pair in state $\tilde\varrho$'' will be entangled with
states of measuring  apparatuses  (and environment) indicating these events.
Then we see, that our total system becomes more and more entangled in a
various possible ways, so that it is rather impossible to perform
the balance of information.

Fortunately, in a recent work Rains \cite{Rains} showed  that any distillation
protocol can be replaced with a deterministic one, achieving the
same distillation rate:
\[
\varrho^{\otimes n}\quad \rightarrow \varrho_{out}\simeq
|\psi_{distilled}\rangle\langle\psi_{distilled}|\otimes \varrho_{rejected}
\]
where $\psi_{distilled}$ is the state of $m$ distilled singlet pairs
while $\varrho_{rejected}$ is the state of the rejected pairs.
In this case the system can be divided into two parts
\be
S=S_{distilled}+S_{rejected}
\label{podzial}
\ee
where $S_{distilled}$ is disentangled with the rest of universe
$S_{rejected}$ is entangled with $M$, hence also with environment $R$.

This possibility of the clear partition into two systems
is crucial for our purposes.
Now the whole balance can be be preformed in this case
as follows. As an input we have the state $\varrho $
with value of asymptotic entanglement of formation $E=E_{F}(\varrho)$.
Because it is mixed we can take its purification
adding come ancilla
which would have the asymptotic entanglement $E'=E+(E'-E)$.
Now we can perform the distillation process,
having no access to the ancilla.
After the process the state of our whole system is still separated
according to the formula (\ref{podzial})
but now the state $S_{rejected}$ involves the degrees of freedom
of the ancilla. The balance of the information can now be easily performed
taking, in particular, into account that distillable entanglement
$E_{D}$ can be interpreted as a useful work (\ref{useful}) $W_{u}$ (Alice
can always teleport through state $|\Psi_{distilled}\rangle \langle \Psi_{distilled} |$
if she wishes). To make the balance fully consistent one should
substract from both input and output data the additional entanglement $E'-E$
coming from extension of the system to the pure state.
As the input physical work (connected with optimal distillation
protocol) {\it is the same} regardless of the value $E'-E$ and
the kind of the ancilla itself, the whole balance is completely consistent.
The input quantities of $E$, $\Delta=(E'-E)$ plus
$W_{p}$ as well as the output ones
$E_D=W_{u}$, $\Delta$, $E_{out}=E(\varrho_{rejected})=E_{bound}$
are depicted on figure Fig. 2b. In particular if we deal with BE states
then the corresponding diagram takes the form of Fig.~2c.

\section{Objectivity of quantum information:
information interpretation of quantum states}

As we have dealt with balance of information
in quantum composite systems it is natural to
ask about objectivity of the entity we qualify.
In this section we discuss that question and
related ones in the context of quantum information
theory and interpretational problems of quantum mechanics.
As one knows the latter defens oneself wery well against commonly
accepted interpretation. In result a number of different interpretations
permanently grows while there is no operational criterions (exept, may be,
Ockham reazor) to eliminate at least some of them.

It is characteristic that despite of dynamical development of interdisciplinary
domain - quantum information there is no, to our knowledge, impact of
the latter
on interpretational problems. In this context a basic question arises:
Does quantum information phenomena provide objective promises for
existence of ``natural'' ontology inherent in quantum formalism?

It is interesting that from among discovered recently quantum effects
just quantum cryptography provides answer ``yes''. To see it clearly,
consider quantum cryptographic protocol. A crucial observation is that
the possibility of secret sharing key  is due to the fact that we
send quantum states {\it themselves} not merely the {\it classical information}
about them! \cite{Doktor}
Clearly, the latter could be cloned by the eavesdropper and it is reason
for which all classical cryptographic schemes are, in principle,
not secure.
Then the use of qubits is {\it crucial} if we would like to take any advantage of the novel
possibilities offered quantum information theory.

Now, as there are experimental implementations of quantum
information protocols  \cite{exp}, it follows
that quantum information is objective and it can provide
natural ontological basis  for interpretation of quantum
mechanics. Then we arrive at important conclusion:
{\it Quantum states carry two complementary kinds
of information: the ``classical'' information involving
quantum measurements and ``quantum'' information
that can not be cloned} \cite{komentarz}.

Note that it is consistent with proposed earlier information intepretation of
the wave function in terms of {\it objective} information content \cite{P89}.
On the other hand it contradicts the Copenhagen interpretation
according to which the wavefunctions have no objective meaning
and only reality is the result of a measurement.
It is remarkable that the above information
interpretation of quantum states is compatible with the
above mentioned unitary information field concept
which rests in the assumption that information is physical
\cite{tata,Landauer,DiVincenzo} and can be defined independently
of probability itself. First axiomatic definition of
classical information ``without probabilities'' was considered by
Ingarden and Urbanik \cite{IngardenU}.
Quantum version of the definition was introduced
by Ingarden and Kossakowski \cite{IngardenK}.
On the other hand Kolmogorow \cite{Kolmogorow}, Solomonoff \cite{S},
Chaitin \cite{C} introduced the concept of classical alghoritmic
information or complexity. Recently the classical alghoritmic
information was incorporated to the definition of the so called
physical entropy  being a constant of ``motion''
under the ``demonic evolution'' \cite{Z,analogy}.

Quite recently alghoritmic information theory
was extended in different ways to quantum states
by Vitanyi \cite{Vita} and Berthiaume {\it et al.} \cite{Berth}.
In fact one can convince oneself that the approaches
\cite{Vita} and \cite{Berth} correspond to the above
complementary kinds of information associated with quantum
state. Indeed, Vitanyi alghoritmic complexity measures
amount of ``classical'' information in bits necessary to approximate
the quantum state. Needless to say, form the point of view of quantum
cryptography such information is useless.
On the other hand the bounded fidelity version of quantum Kolmogorow
complexity measures amount quantum information in a qubit string
and it is closely related to quantum compression theory
\cite{Schum,JS,JH3}.
\section{Summary}
In conclusion we have developed the  entanglement -- energy  analogy
based on some natural postulates:
(i) entanglement is a form of quantum information being counterpart
of internal energy,
(ii) the process of sending qubits
as a counterpart of work.
We also assume that the evolution of the quantum system is unitary.

Basing on the above postulates we have considered the balance of
quantum information for bipartite quantum communication systems
i. e. the systems composed of two spatially separated
laboratories endowed with classical informational
channel plus local quantum operations.
Wa have introduced the notion of informational content of quantum
state being a difference of maximal possible von
Neumann entropy and the actual one. Then we have defined
physical work as a number of qubits physically
sent form Alice to Bob. We have also defined logical work as
as increase of the informational content of Bob state.
To have a proper description of quantum communication processes
we have also introduced a notion of useful logical work
as amount of qubits transmitted without decoherence.

Those tools have allowed us to perform the
detailed balances of quantum information in two
important processes of quantum communication:
quantum teleportation and distillation of quantum noisy entanglement.
In particular we have discuss the question of balance of quantum
information for open systems.
In the context of balance scheme and related notions
we conclude that the irreversibility
connected with existence of bound entanglement
can be viewed as an analogue to irreversible thermodynamical processes
where only the free energy (which is not equal to the
total energy) can be converted to useful work.
This allows us to interprete the equation
for entanglement of formation as quantum information counterpart
of the thermodynamical Gibbs-Hemholtz equation.

Finally we discuss the objectivity of quantum information
in general context of some recent achievments of quantum information theory
including quantum cryptography and recent propositions
of classical and quantum alghorytmic information.
This leads us to the conclusion that quantum states
reflect properties of quantum information as objective
entity involving ``classical'' and ``quantum'' components
which correspond to recently introduced
``classical'' and ``quantum'' alghoritmic complexities.
So the balance performed in the present paper
concerns objective quantities rather than purely formal objects.
We hope that the present informational approach
to bipartite quantum communication systems, when suitably developed,
may lead to deeper understanding of quantum information processing
domain.

M. H. and P. H. thank Chris Fuchs and Pawe\l{} Masiak
for discussions on quantum information. Part of this work was
made during ESF-Newton workshop (Cambridge 1999).
The work is supported by Polish Committee for Scientific
Research, contract No. 2 P03B 103 16.

\begin{figure}
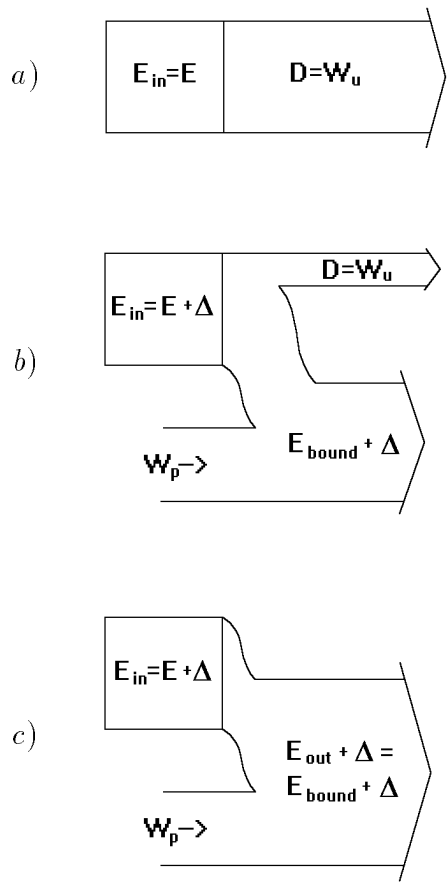

\caption[Balances of information]
{The diagram illustrating balance of quantum information
in entanglement distillation process for:
(a) pure states case, (b) general case, (c) bound entangled states case.}
\end{figure}

\begin{references}
\bibitem{footTot} Alternative name is the ``total'' entanglement as it was
used originally in \cite{APS}.
\bibitem{foot3}
Here $E_{f}$ is entanglement of formation defined in Ref. \cite{huge}.
\end{references}
\end{document}